\documentstyle[12pt,dina4,psfig]{article}
\newcommand{\bc}{\begin{center}}
\newcommand{\ec}{\end{center}}
\newcommand{\bt}{\begin{tabular}}
\newcommand{\et}{\end{tabular}}

\newcommand{\be}{\begin{equation}}
\newcommand{\ee}{\end{equation}}
\date{}
\begin{document}
\title{Medium effects in the production and decay of vector mesons
in pion-nucleus reactions\footnote{Supported by DFG}}

\author{ Ye.S. Golubeva$^1$, L.A. Kondratyuk$^2$ and W. Cassing$^3$}
\maketitle

\bc$^1$ Institute of Nuclear Research, 117312 Moscow, Russia
\\

$^2$ Institute of
Theoretical and Experimental Physics, 117259
Moscow, Russia \\

$^3$ Institute  for Theoretical Physics, University of
Giessen, D-35392 Giessen, Germany \\

\ec

\begin{abstract}
The $\omega$-, $\rho$- and $\phi$-resonance production and their dileptonic
decay in  $\pi^-   A$ reactions at 1.1 - 1.7 GeV/c are calculated within the
intranuclear cascade (INC) approach. The invariant mass
distribution of the dilepton pair for each resonance can be decomposed in
two components which correspond to their decay
'inside' the target nucleus and in the vacuum, respectively.
The first components are strongly
distorted by the nuclear medium due to resonance-nucleon scattering and
a possible mass shift at finite baryon density. These medium modifications
are compared to background sources in the dilepton spectrum from $\pi N$
bremsstrahlung as well as the Dalitz decays of $\omega$ and $\eta$ mesons
produced in the reaction. Detailed predictions for $\pi^-  Pb$ reactions
at 1.3 and 1.7 GeV/c are made within several momentum bins for the lepton
pair.
\end{abstract}

\vspace{2cm}


\newpage

\section{Introduction}

The question about the properties of hadronic resonances in the nuclear 
medium has received a vivid attention during the last years (cf. Refs. 
\cite{1,2,Klingl,3,4}). Here, QCD inspired effective Lagrangian models 
\cite{1,2,Klingl} or approaches based on QCD sum rules \cite{3,4} 
predict that the masses of the vector mesons $\rho$, $\omega$ and 
$\phi$ should decrease with the nuclear density. Furthermore, along 
with a dropping mass the phase space for the resonance decay also 
decreases which results in a modification of the resonance width in 
matter.  On the other hand, due to collisional broadening - which 
depends on the nuclear density and the resonance-nucleon interaction 
cross section ( cf. Refs. \cite{5,6}) - the resonance width will 
increase again.

The in-medium properties of vector mesons have been addressed 
experimentally so far by dilepton measurements at the SPS, both for 
proton-nucleus and nucleus-nucleus collisions 
\cite{CERES,HELIOS,HELI,Drees}.  As proposed by Li {\it et al.} 
\cite{Li}, the enhancement in $S + Au$ reactions compared to $p + Au$ 
collisions in the invariant mass range $0.3 \leq M \leq 0.7$ GeV might 
be due to a shift of the $\rho$ meson mass. The microscopic transport 
studies in Refs. \cite{11,12,brat} for these systems point in the same 
direction, however, also more conventional selfenergy effects cannot be 
ruled out at the present stage \cite{11,Rapp,Wamb,Friman}.  It is 
therefore necessary to have independent information on the vector meson 
proporties from reactions, where the dynamical picture is more 
transparent, i.e. in pion-nucleus collisions. Here, especially the 
$\omega$ meson can be produced with low momenta in the laboratory 
system, such that a substantial fraction of them will also decay inside 
a heavy nucleus \cite{Schoen,CGIK}.  The same holds for the $\phi$ 
meson, however, its vacuum decay will still dominate except for very 
low momenta in the laboratory.

The mass distributions of the vector mesons in the latter case are 
expected  to have a two component structure \cite{6} in the dilepton 
invariant  mass spectrum: the first component corresponds to resonances 
decaying in the vacuum, thus showing the free spectral function which 
is very narrow in case of the $\omega$ or $\phi$ meson; the second 
(broader) component then corresponds to the resonance decay inside the 
nucleus. We will use that (in first order) the in-medium resonance can 
also be described by a Breit-Wigner formula with a mass and width 
distorted by the nuclear environment.

In this paper we will carry out microscopic calculations for  the 
production and dileptonic decay of $\omega$, $\rho$ and $\phi$ 
resonances in $\pi^- A$ collisions at pion momenta of 1.1 - 1.7 GeV/c 
available at GSI in the near future. The calculations are performed  
within the framework of the intranuclear cascade model (INC) \cite{7} 
which was extended earlier \cite{8,9} to account for the in-medium 
resonance decays. First calculations for the shapes of the $\omega$ and 
$\rho$ peaks in proton (antiproton) induced reactions $p A\to VX\to 
e^+e^-X$ ($\bar p A\to VX\to e^+e^-X$) have been reported in Refs. 
\cite{8,9}.  Here we consider explicitly $\pi^-$ induced reactions and 
also compute the background sources in the dilepton spectrum from 
pion-nucleon bremsstrahlung as well as the Dalitz decays $\omega \to 
\pi^0e^+e^-$ and $\eta \to \gamma e^+e^-$ following Refs. 
\cite{11,12,CGIK,10}.

Our work is organized as follows: In Section 2 we will briefly derive 
the two-component structure of the resonance decays and provide a 
description of the intranuclear cascade model (INC) including the 
resonance cross sections employed. In Section 3 we analyse in detail 
the production and decay of $\omega$ mesons as a function of their 
laboratory momentum, the nuclear target as well as the initial pion 
laboratory momentum. Section 4 is devoted to a study of $\phi$ meson 
production and decay in $\pi^-  Pb$ reactions at 1.7 GeV/c while 
section 5 concludes our study with a summary and discussion of open 
problems.

\section{Theoretical framework}

\subsection{The two-component structure of the resonance decay}
\indent
     As it was shown in Ref. \cite{6} the amplitude describing
 the production of a resonance in a finite nucleus is a superposition
 of two contributions arising from decays of the resonance 'inside'
 and 'outside' the nucleus. In Ref. \cite{6} the main emphasis was put on
 the derivation of the Green function which describes the
 propagation of the hadronic resonances in the nuclear medium.
 Whereas the latter work was performed
 in the  framework of multiple scattering theory, here we
 present an equivalent derivation using the optical potential approach
 which  was applied successfully
 to the description of  pion-nucleus interactions before (cf.  \cite{ericson}).

 We start from the relativistic form of the wave
 equation which describes the propagation of a resonance $R$ in the
 nuclear medium
\be
\label{waveeq}
(-{\vec{\nabla}^2} + M^2_R - iM_R\Gamma_R + U(\vec{r}))\Psi(\vec{r})=
E^2\Psi(\vec{r}),
\ee
where  $E^2={\vec{p}^2}+ M^2_R$; $\vec{p},M_R$ and $\Gamma_R$ are
the momentum,
mass and width of the resonance, respectively.
The optical potential then is defined as
\be
\label{optpot}
 U(\vec{r})= - 4\pi f(0)\rho_A(\vec{r})
\ee
where $f(0)$ is the forward $RN$-scattering amplitude and $\rho_A$ is
the nuclear density.

It is useful to rewrite Eq. (\ref{waveeq}) in the form

\be
\label{waveeq2}
({\vec{\nabla}^2} + \vec{p}^2)\Psi(\vec{r})=(U(\vec{r}) -
 \Delta)\Psi(\vec{r}),
\ee
where
\be
\label{delta}
 \Delta = P^2- M^2_R - iM_R \Gamma_R
\ee
is the inverse resonance propagator.
The four-momentum P in (\ref{delta}) can be defined through the four-momenta
of the resonance decay products,
\be
P=p_1+p_2+...
\ee

Let us consider the case of a fast resonance, i.e. $pR_A \gg 1$, where
$R_A$ is the nuclear radius. In case of a Pb-target this implies
$p \gg$ 30 MeV/c.  For such a case the eikonal approximation
can be used and  the Green function describing the propagation of the
resonance from the point $\vec{r}=(\vec{b},z)$ to $\vec{r}'=(\vec{b}',z')$
can be written as
$$
G_p(\vec{b}',z';\vec{b},z)=\frac{1}{2ip} exp
\{i\int^{z'}_z[p+{\frac{1}{2p}(\Delta +4\pi f(0)\rho_A(\vec{b}},\zeta))]
d\zeta\}
$$
\be
\label{Green}
\times \delta(\vec{b}-\vec{b}')\theta(z'-z),
\ee
where the $z$-axis is directed along the resonance momentum $\vec{p}$,
while $\vec{b}$ is the impact parameter.

We will consider here pion energies slightly above threshold, such that
the vector mesons  $\rho, \omega$ and $\phi$ can be produced only
in the first hard pion-nucleon
collision. Being produced in the point   $\vec{r}=(\vec{b},z)$ inside
the nucleus the vector meson then propagates to
the point  $\vec{r}'=(\vec{b}',z')$ , where it decays into a dilepton pair
with the total momentum $\vec{P}$ and invariant mass $\vec{P^2}$.
The corresponding amplitude can be written as
\be
\label{both}
M(\vec{P},P^2;\vec{b},z)=N f_{\pi^-N_j \to RX_i}
\int d^2\vec{b'} dz' \{exp(-i \vec{p} \vec{r} ') G_p(\vec{r} '-\vec{r})
exp(i \vec{p} \vec{r})\} f_{R \to e^+e^-},
\ee
where $f_{\pi^-N_j \to RX_i}$ is the production amplitude,
$X_i = N, \Delta,...$, $f_{R \to e^+e^-}$ is the decay amplitude
and $N$ a normalization factor which includes also the
shadowing effect.

The point  $\vec{r} '=(\vec{b}',z')$  of the resonance decay
can be inside or outside the nucleus. Accordingly the exponent of the Green
function (\ref{Green})  can be
separated into the contributions from the two regions

1) $z\leq z'\leq z_s= \sqrt{R_A^2-\vec{b}^2}$ and

2) $z'\geq z_s=\sqrt{R_A^2-\vec{b}^2}.$

\vspace{0.5cm}
\noindent
When the resonance decays inside the nucleus of radius $R_A$, 
only the first part contributes and the inverse resonance 
propagator has the form
\be
\Delta^*=\Delta+4\pi  f(0)\rho_0= P^2-M^{*2}_R+iM_R^*\Gamma_R^*
\ee
where
\be
M_R^{*2}=M^2_R-4\pi Re{f(0)}\rho_A,
\ee
\be
M_R^*\Gamma^*_R=M_R\Gamma_R+4\pi Im{f(0)}\rho_A.
\ee
If the resonance decays outside the nucleus, both regions contribute
and the amplitude for lepton pair production can be written as
\be
\label{both1}
M^i_j(\vec{P}, P^2;\vec{b},z) = N \  f_{\pi^-N_j \to RX^i}
\{A_{in}(\vec{P},P^2; \vec{b},z)+
A_{out}(\vec{P}, P^2; \vec{b},z)\} f_{R \to e^+e^-}.
\ee
In Eq. (\ref{both1}) the contributions from the first and second part can be
expressed as
\be
A_{in}(\vec{P}, P^2; \vec{b},
z)=\frac{1-exp[i(\Delta^*/2k)(z_s-z)]}{\Delta^*},
\ee
and
\be
A_{out}=\frac{exp[i(\Delta^*/2k)(z_s-z)]}{\Delta}.
\ee
Finally, the dilepton invariant mass spectrum is given by
the following expression:
\be
\label{sig}
\frac{d\sigma}{dM} = \sum_{i,j} \int d^2b dz \ \rho(\vec{b},z) \
|M^i_j(\vec{P},\vec{P}^2,\vec{b},z)|^2 ,
\ee
where the sum is taken over all nucleons in the target and all
production channels, respectively.

When the resonance decays inside the nucleus, its explicit form is
described (in a first approximation) by a Breit--Wigner formula
\be
\label{BW}
F(M)= \frac{1}{2 \pi} \frac{\Gamma^*_R}{(M-M^*_R)^2 + \Gamma^{*2}_R/4}
\ee
that contains the effects of collisional broadening,
\be
\label{gammas}
\Gamma^*_R=\Gamma_R+\delta \Gamma ,
\ee
where
\be
\label{dgamma}
\delta\Gamma = \gamma v\sigma_{(RN)}\rho_A,
\ee
and a shift of the meson mass
\be
\label{mstar}
M^{*}_R=M_R+\delta M_R ,
\ee
where
\be
\label{dmstar}
\delta M_R=-\gamma v \sigma_{(RN)}\rho_A\alpha.
\ee
In Eqs. (\ref{dgamma}) and (\ref{dmstar}) $v$ is the resonance
velocity with respect to the target at rest, $\gamma$ is the
associated Lorentz factor, $\rho_A$ is the nuclear density, $\sigma_{(RN)}$ is
the resonance--nucleon total cross section and $\alpha = (Re f(0))/(Im f(0)).$

If the ratio $\alpha$ is small - which is actually the case for the 
reactions considered because many reaction channels are open - the 
broadening of the resonance will be the main effect. The sign of the 
mass shift depends on the sign of the real part of the forward $RN$ 
scattering amplitude which, in principle, also depends on the momentum 
of the resonance. For example, at low energy various authors  
\cite{1,2,3,4}  predict a decreasing mass of the vector mesons $\rho$, 
$\omega$ and $\phi$ with the nucleon density, whereas Eletsky and Ioffe 
have argued  recently \cite{13} that the $\rho$ should become heavier 
in nuclear matter at momenta of  2-7 GeV/c. We will not address this 
question further since we will concentrate essentially on vector mesons 
with low momenta relative to the nuclear target and model the mass 
shift $\delta M_R$ independently.

\subsection{The intranuclear cascade model}
In the following we consider the reaction $\pi^- A\to VX\to e^+e^-X$ 
for different targets at pion  momenta from 1.1 up to  1.7 GeV/c.  The 
yields of vector mesons ($\omega$,  $\rho$ and $\phi$) are calculated 
 within the framework of the intranuclear cascade model (INC) developed  
in Ref. \cite{7,15}.  In the INC the linearized kinetic equation for 
the many-body distribution function - describing the hadron transport 
in nuclear matter \cite{bun} - is solved numerically by  assuming that 
during the evolution of the cascade the properties of the target 
nucleus remain unchanged. This implies that the number of cascade 
particles $N_c$ is much less than the number of nucleons $A_t$ in the 
target nucleus, i.e. $N_c \ll A_t$. Since the nucleus is a finite open 
system with a relatively small number of nucleons, the latter condition 
is violated for $\pi N$ interactions at energies $E_{\pi} >$ 3 - 5 GeV 
\cite{bor} when multiple pion production becomes dominant.

Within the INC approach the target nucleus is regarded as a mixture of 
degenerate neutron and proton Fermi gases in a spherical potential well 
of diffuse surface. The momentum distribution of the nucleons is 
treated in the local density approximation for a Fermi gas.  
Restricting ourselves to pion momenta below 1.7 GeV/c we have to take 
into account the following elementary processes:
\be
\label{1}
\pi^- p\to \omega n,
\ee
\be
\label{2}
\pi^-N \to \omega \pi N
\ee
\be
\label{3}
\pi^- p \to \rho n,
\ee
\be
\label{4}
\pi^-N \to \rho \pi N
\ee
\be
\label{5}
\pi^- p\to \phi n .
\ee
For  the process (\ref{1}) we use a parametrization of the experimental data
from Ref. \cite{15,Cugnon}:
\be
\sigma(\pi^- p \to \omega n)= C \frac{P_{\pi N}-P_{\omega}^0}{P_{\pi 
N}^{\alpha}-d},
\ee
where $P_{\pi N}$ is the relative momentum (in GeV/c) of the pion-nucleon
pair while $P_{\omega}^0$ = 1.095 GeV/c is
the threshold value. The parameters $C = $ 13.76 mb (GeV/c)$^{\alpha -1}$,
$\alpha$=3.33 and $d$=1.07 (GeV/c)$^{\alpha}$ describe satisfactorily the data
on the energy-dependent cross
section (\ref{1}) in the near-threshold energy region.

For the inclusive $\omega$ production (\ref{2}) we use the
parametrization from Ref. \cite{14}:
\be
\label{incl}
     \sigma(\pi^- p \to \omega \pi N) = a(x-1)^b x^{-c},
\ee
where the scaling variable is defined as $x=s/s_{th}$; for $\omega$-
production we have $s_{th}$ = 2.958  GeV$^2$, $a=4.8 mb$, $b=1.47$, $c=1.26$.
Furthermore, we use
\be
\label{incl2}
\sigma_{\pi^- n \to \omega \pi N} = 3 \sigma_{\pi^- p \to \omega \pi N}
\ee
for the isospin dependence in the entrance channel according to a $\Delta$
intermediate state.

For $\rho^0$-meson
production (\ref{3}), (\ref{4}) we use the same cross sections as for
$\omega$-mesons; this holds experimentally within 20\%.
Since we will calculate $\phi$ meson production at a pion momentum of
1.7 GeV/c,  only channel (\ref{5}) contributes to $\phi$ meson production.
In our calculations we use the parametrization of the
cross section  from Ref. \cite{14} for the exclusive channel, i.e.
\be
\sigma_{\pi^- p \rightarrow \phi N} (s) = A \frac{\Gamma^2}{(\sqrt{s} - M_R)^2
+ \Gamma^2/4} \ \frac{\pi^2 |p_\phi|}{4 p_{\pi}^2 s}
\ee
with $A$ = 0.00588 mb GeV$^3$, $\Gamma$ = 0.99 GeV and $M_R$ = 1.8 GeV, where
$p_\pi$ and $p_\phi$ denote the pion and $\phi$ momentum in the cms,
respectively. The angular distribution for the products of all reactions
(\ref{1}) - (\ref{5}) are considered
to be isotropic in the pion-nucleon c.m.s. since we operate
close to threshold energies.

Vector mesons in this kinematical range are produced only in the first 'hard'
interaction. Since the cross sections for their production are small
( about 2 mb  for $\omega$ and $\rho$ and 20 $\mu b$ for $\phi$), we use
the weight - function method to calculate  the
vector meson production and decay in nuclei, i.e. each vector meson carries
a weight
\be
\label{weight}
W_i = \frac{\sigma_{\pi^- N \to R+X}(\sqrt{s})}{\sigma_{\pi^- N}^{tot} 
(\sqrt{s})},
\ee
where $\sigma_{\pi^- N}^{tot}$ is the total $\pi N$ cross section at 
invariant energy $\sqrt{s}$.

The propagation of vector mesons that are produced in the elementary 
subprocesses is described in the same manner as in our previous works 
\cite{7,15}. Furthermore, the $\omega$-, $\rho$- and $\phi$-mesons may 
interact with nucleons or decay inside the nucleus into mesons and 
dileptons.  The competition between their decay to mesons and their 
interaction with a nucleon is determined by the following expression 
for the mean free path:
\be
\label{path}
  1/\lambda=1/\lambda_{dec}+1/\lambda_{int}  ,
\ee
were $\lambda_{int}=(\rho_{A}\sigma^{tot}_{(VN)})^{-1}$,
 $\lambda_{dec}=\gamma\beta /(\Gamma_R)$,
$\rho_{A}$ is the nuclear density,
$\sigma^{tot}_{(RN)}$ is the total cross section for the  meson-nucleon
interaction, $\gamma=(1-\beta^2)^{-1/2}$
 is the Lorentz factor, $\beta$ the particle velocity
in units of c, while $\Gamma_R$ is the vector meson vacuum width.

The "fate" of $\omega$- and $\phi$- mesons in the intranuclear cascade  is
determined by the total cross sections $(\sigma^{tot}_{\omega N})^{-1}$,
$(\sigma^{tot}_{\rho N})^{-1}$, $(\sigma^{tot}_{\phi N})^{-1}$
and the partial cross sections for the
following interactions with nucleons:
\be
\label{c1}
  \omega N \to \omega N, \  \omega N \to \pi N, \  \omega N \to \pi \pi N,
\ee
\be
\label{c2}
  \rho N \to \rho N, \rho N \to \pi N, \rho N \to \pi \pi N,
\ee
\be
\label{c3}
  \phi N \to \phi N, \phi N \to \pi N, \phi N \to \pi \pi N.
\ee
For these reactions no experimental data are directly available and we
have to introduce 'plausible' parametrizations.
As a model for the $\omega$- and $\phi-N$ cross sections  we use for the 
total cross section
\be
\label{c4}
\sigma_{\omega N}^{tot} (p_{lab}) = A + \frac{B}{p_{lab}}
\ee
with A = 11 mb and B = 9 mb GeV/c; in case of the $\phi$ meson we adopt
A = 5 mb and B = 4.5 mb GeV/c. The form (\ref{c4}) guarantees that the
collisional width $\delta \Gamma_R$ (17) does not diverge, but becomes
constant for $p_{lab} \to $ 0. For the elastic $\omega N$ cross section
we, furthermore, adopt
\be
\label{c5}
\sigma^{el}_{\omega N} (p_{lab}) = A \frac{1}{1 + a p_{lab}}
\ee
with A = 20 mb and $a$ = 1 GeV$^{-1}$c. In case of elastic $\phi N$
reactions we use A = 10 mb and also $a$ = 1 GeV$^{-1}c$.

For the $\rho$-N total and elastic cross sections we adopt the results 
of Ref.  \cite{Sibirtsev}, which have been determined from experimental 
data for the $\pi N \to \rho N$ exclusive cross section in a 
meson-nucleon-resonance model using the experimental branching ratios 
for the resonances involved \cite{PDB}.  The channels $\omega N 
\rightarrow \pi N, \rho N \rightarrow \pi N$ are determined via 
detailed balance from the inverse reactions (26).

In order to explore the observable consequences of vector meson mass 
shifts at finite nuclear density the in-medium vector meson masses are 
modelled according to Hatsuda and Lee \cite{3} as
\be
\label{Brown}
M^*_R=M_R(1 - \alpha \rho_A(r) / \rho_0),
\ee
where $\rho_A (r)$ is the nuclear density at the
resonance decay, $\rho_0 = 0.16 fm^{-3}$, and $\alpha = 0.18$ for the $\rho$
and $\omega$ while $\alpha = 0.025$ is taken for the $\phi$ meson as in
Ref. \cite{Ehehalt}.

The decay of the resonances to $e^+e^-$ with their actual spectral 
shape is performed as in  Refs. \cite{8,9}:  when the resonance decays 
into dileptons inside the nucleus its mass is generated  according to a 
Breit--Wigner distribution with average mass $M^*_R$  (\ref{Brown}) and 
$\Gamma^*_R=\Gamma_R+\delta \Gamma_R$ (\ref{gammas}), where the 
collisional broadening  and the mass shift are calculated according to 
the local nuclear density. Its decay to dileptons is recorded as a 
function of the corresponding invariant mass bin and the local nucleon 
density $\rho_A$.  If the resonance leaves the nucleus, its spectral 
function automatically coincides with the free distribution because 
$\delta M_R$ and $\delta \Gamma_R$ are zero in this case.

\section{Production and decay of $\omega$ mesons}
Following the suggestion by Sch\"on et al. \cite{Schoen} we start with 
the reaction $\pi^- + ^{208}Pb$ at 1.3 GeV/c and present the calculated 
longitudinal ($P_z$) and transverse ($P_T$) momentum distributions of 
$\omega$ mesons in Fig. 1 which decay inside (solid histograms) and 
outside (dashed histograms) of the Pb-nucleus. Since the $\omega$ 
decays have been recorded as a function of the nucleon density, the 
'inside' component is defined as those $\omega$ mesons which decay at 
densities $\rho \geq$ 0.03 $\rho_0$. As one might expect due to 
kinematical reasons the longitudinal momentum distribution of $\omega$ 
mesons for the 'inside' component is shifted to lower momenta $P_z$ 
whereas fast $\omega$'s predominantly decay in the vacuum.  A similar 
correlation also holds for the transverse momentum distribution though 
it is not as pronounced as for the longitudinal momentum distribution. 
Thus it is clear that in order to study in-medium decays of $\omega$ 
mesons cuts for low $P_z$ and $P_T$ are favorable.

\subsection{Dilepton decay of $\omega$ mesons}
Including the mass shift of the $\omega$'s as well as collisional 
broadening (as described in Section 2) we show in Fig. 2 the inclusive 
differential cross section of the $e^+e^-$ pairs from direct $\omega$ 
decays for different cuts in $P_z$ and $P_T$.  In the lowest momentum 
interval ($P_z \leq$ 0.25 GeV, $P_T \leq 0.25$ GeV) one clearly 
observes a two peak structure corresponding to the in-medium and vacuum 
decays, respectively. In order to quantify the ratio from both 
components we have introduced a cut in invariant mass at $M_c$ = 0.725 
GeV and integrated the dilepton yield below and above $M_c$. The quantity
\begin{equation}
\label{R}
R = \frac{\int_0^{M_c} dM N_{e^+e^-} (M)}{\int_{M_c}^\infty dM N_{e^+e^-} (M)}
\end{equation}
thus provides a measure for the in-medium $\omega$ decay relative to 
the vacuum decay. Its actual values for the various momentum cuts are 
given in Fig. 2 and decrease from R = 1.16 (for the lowest momentum 
bin) with increasing total momentum of the dilepton pair to R=0.21 (for 
the highest momentum bin).  Thus in case of a large momentum acceptance 
of the dilepton spectrometer (HADES \cite{HADES}) one can use different 
cuts in longitudinal and transverse momentum to explore the in-medium 
properties of the $\omega$ meson as a function of its momentum with 
respect to the target nucleus at rest, too.

Since in light nuclei the vacuum decay of the $\omega$ will dominate, 
one has to investigate the target mass dependence of the in-medium 
component, respectively. The results of our calculations at a pion 
laboratory momentum of 1.3 GeV/c are displayed in Fig. 3 for $^{208}Pb, 
^{90}Zr, ^{40}Ca$ and $^{12}C$ in the lowest momentum bin ($P_z \leq$ 
0.25 GeV/c, $P_T \leq$ 0.25 GeV/c) for the dilepton pair. The ratio R 
(\ref{R}) here decreases from R = 1.16 to R = 0.44 when going from the 
heavy (Pb) to the light target (C).  In case of $^{12}C$ no explicit 
in-medium peak is visible anymore even for the lowest momentum cut; 
only a low mass tail of the pronounced peak for the vacuum decay 
appears.  This explicit mass dependence can also be exploited 
experimentally to prove or disprove in-medium modifications of the 
$\omega$ meson by directly comparing dilepton spectra from light and 
heavy targets (cf. Section 3.3).

A further question is related to the dependence of the in-medium 
component as a function of the pion laboratory momentum. One expects 
the $\omega$ production cross section to increase with the pion 
momentum, however, their average momentum distribution will be shifted 
to higher momenta, too, such that the in-medium decay component might 
be reduced at higher energy. We have thus performed calculations for 
$\pi^- + Pb$ collisions for pion momenta of 1.1, 1.3, 1.5, and 1.7 
GeV/c, respectively.  The results of our computations for $\pi^- + Pb$ 
reactions are given in Fig. 4 for the lowest momentum bin ($P_z \leq$ 
0.25 GeV/c, $P_T \leq$ 0.25 GeV/c) of the dilepton pair. The ratio R 
here (within the numerical accuracy) is roughly constant from 1.3 - 1.7 
GeV/c while the cross section increases by about 50\%. The largest 
signal R = 1.44 we obtain at 1.1 GeV/c, however, here the cross section 
is already down by about a factor of 2 as compared to a beam momentum 
of 1.3 GeV/c. Thus our analysis favors laboratory momenta of about 1.3 
GeV/c for the study of the in-medium properties of the $\omega$ meson, 
which is in line with the suggestion by Sch\"on et al. \cite{Schoen}.

\subsection{Variations of the $\omega$ in-medium properties}
As discussed in Section 2 the properties of the $\omega$ meson due to 
collisional broadening and a mass shift at finite nucleon density are 
presently not well known. We thus have to explore different medium 
effects in order to see if the in-medium $\omega$ signal will survive.  
In this respect we first (artificially) increase the elastic $\omega$ N 
cross section (35) by a factor of 2, but keep the $\omega$ mass shift 
fixed by Eq. (\ref{Brown}). The respective $\omega$ momentum 
distributions in the laboratory are displayed in Fig. 5 for the 
enhanced elastic cross section (full histograms) and the expression 
(\ref{c5}) (dashed histograms).  Here the $\omega$ momentum 
distribution for the inside decay is displayed in the upper part of 
Fig. 5 while the distribution for the vacuum decay component is shown 
in the lower part of the figure. We find that a larger elastic cross 
section leads to a more efficient stopping of the $\omega$ mesons in 
the nuclear target as seen by the shift of the momentum distributions 
to lower momenta. As a consequence more $\omega$'s decay inside the 
target with increasing elastic scattering cross section.

This is shown quantitatively in Fig. 6 for the reaction $\pi^-  Pb$ at 
1.3 GeV/c for the lowest momentum bin ($P_z \leq $ 0.25 GeV/c, $P_T 
\leq $ 0.25 GeV/c).  With increasing elastic cross section the signal 
shape in invariant mass (histogram a) versus histogram b)) is not 
changed very much, however, the inclusive cross section in the lowest 
momentum bin increases significantly; the ratio R $\approx$ 1.16 for 
both cases is practically the same.  Thus by measuring the dileptons 
from the $\omega$ decay at different momentum bins one is able to 
extract information on the in-medium scattering cross section of 
$\omega$'s with nucleons as well.

Another problem is related with the actual mass shift of the $\omega$ 
at finite density. In Eq. (\ref{Brown}) we have adopted a coefficient 
of 0.18 as suggested by QCD sumrules \cite{3}. The sumrule studies are 
performed for $\omega$ meson at rest in the nucleus, however, the 
$\omega$ polarization or selfenergy might well be a function of 
momentum (as suggested by the studies in Refs. \cite{Wamb,Friman} in 
case of the $\rho$ meson) such that the actual mass shift seen in 
$\pi^-  Pb$ reactions might be different.  A clear signal is still 
expected if the $\omega$ mass shift is larger than that assumed in Eq. 
(\ref{Brown}); we thus investigate the question if it still can be 
extracted from dilepton spectra when it reduces to 50\%. The numerical 
results are shown for this case in Fig. 7 (solid histogram) for the 
reaction $\pi^- + Pb$ at 1.3 GeV/c for the lowest momentum bin ($P_z 
\leq$ 0.25 GeV/c, $P_T \leq$ 0.25 GeV/c) in comparison to the result 
according to Eq. (\ref{Brown}) (dashed histogram).  With decreasing 
mass shift of the $\omega$ meson its in-medium decay peak shifts closer 
to the vacuum decay peak, however, with an experimental resolution 
$\Delta M \leq$ 10 MeV it should still be visible in the experimental 
spectrum at least for the lowest momentum bin.

\subsection{Background processes}
A first calculation for the background processes at invariant masses 
above 0.2 GeV for the reaction $\pi^- Pb$ at 1.3 GeV/c has been given 
in Ref.  \cite{CGIK} integrated over all momenta of the dilepton pair. 
In the latter work it was found that the background for invariant 
masses above 0.65 GeV dominantly stems from the $\rho^0$ decay with a 
small contribution from $\pi^- N$ bremsstrahlung, while the resonance 
($\Delta$) Dalitz decays are negligible as well as proton-neutron 
bremsstrahlung.  Here we repeat the studies of Ref. \cite{CGIK} using 
the same formfactors for the $\eta$ and $\omega$ Dalitz decays as well 
as for pion-nucleon bremsstrahlung, however, differentiate with respect 
to momentum bins. The results of our calculations are shown in Fig. 8 
for the reaction $\pi^-  Pb$ at 1.3 GeV/c for the inclusive dilepton 
yield integrated over all momenta (l.h.s.) as well as for the lowest 
momentum bin ($P_z \leq $ 0.25 GeV/c, $P_T \leq$ 0.25 GeV/c) (r.h.s.). 
We find the relative background contributions above 0.65 GeV in the low 
momentum bin to be in the same order of magnitude as for the momentum 
integrated spectrum, whereas the inclusive cross section drops by about 
a factor of 15 for the lowest momentum bin.

For experimental purposes we, furthermore, show the results of our 
calculation including the background processes discussed above for the 
reaction $\pi^-$  $^{12}C$ (dashed histograms) and $\pi^-$  $ ^{208}Pb$ 
(solid histograms) at 1.3 GeV/c in Fig. 9 integrated over all momenta 
(l.h.s.) as well as for the lowest momentum bin ($P_z \leq$ 0.25 GeV/c, 
$P_T \leq$ 0.25 GeV/c) (r.h.s.), respectively. In order to allow for a 
direct comparison, both systems have been normalized to the same 
differential cross section in the $\omega$ vacuum decay peaks. The 
relative enhancement at invariant masses of 0.65 GeV for the Pb-target 
is quite pronounced for the lowest momentum bin, however, survives also 
in the momentum integrated spectra (upper part). Thus the in-medium 
$\omega$ meson properties can be extracted experimentally by comparing 
directly the dilepton yield from light and heavy targets especially for 
low momentum cuts. We note that the relative dilepton yield at 
invariant masses $M \leq$ 0.65 GeV is smaller for $^{12}C$ as for $Pb$ 
because we have normalized to the free $\omega$ decay peak which is 
more pronounced for $^{12}C$ since most of the $\omega$-mesons decay in 
the vacuum here.

\section{$\phi$ meson production and decay}
Apart from the $\omega$ meson the in-medium properties of the $\phi$ 
meson can be studied as well by $\pi^- A$  reactions. Since for pion 
momenta of 1.3 GeV/c the $\phi$ cross section is rather low even for 
Pb-targets, we perform our analysis at a laboratory momentum of 1.7 
GeV/c.  The distribution of $\phi$ mesons in longitudinal momentum 
($P_z$) and transverse momentum ($P_t$) are displayed in Fig. 10 for 
the 'inside' and 'outside' component, respectively. Due to the longer 
lifetime of the $\phi$ mesons (as compared to $\omega$ mesons) and 
lower in-medium scattering cross sections \cite{Ko97}, the 'outside' 
momentum distribution is considerably larger than the 'inside' momentum 
distribution. The in-medium properties of the $\phi$ will thus be 
harder to observe.

In Fig. 11 we show the invariant mass distribution of dilepton pairs 
from the background processes as well as $\omega, \rho^0$ and $\phi$ 
decays including the mass shift (\ref{Brown}) for $\omega$ and $\rho$ 
as well as collisional broadening for all mesons, however, discarding a 
mass shift of the $\phi$ meson. The momentum integrated mass spectra 
for $\pi^-  Pb$ at 1.7 GeV/c are shown in the l.h.s. of Fig. 11 whereas 
a low momentum cut ($P_z \leq$ 0.25 GeV/c, $P_T \leq$ 0.25 GeV/c) has 
been applied in the r.h.s. of Fig. 11. The $\phi$ peak clearly emerges 
out of the background from the $\rho$ decay even for the integrated 
spectra, while the $\phi$ decay is even more pronounced in the low 
momentum bin.  Thus the $\phi$ meson can clearly be studied 
experimentally if the mass resolution $\Delta M$ is 10 MeV or less.

We now address the question of in-medium effects on the $\phi$ meson 
again for the system $\pi^-  Pb$ at 1.7 GeV/c. The results of our 
calculations are displayed in Fig. 12 for a low momentum cut ($P_z 
\leq$ 0.25 GeV/c, $P_T \leq$ 0.25 GeV/c).  The full histogram shows the 
dilepton spectra without a mass shift for $\phi$ (cf. Fig. 11) while 
the dashed histogram includes a $\phi$ mass shift according to 
(\ref{Brown}) with a 2.5\% reduction at $\rho_0$ according to \cite{3} 
leading to an in-medium peak shifted by about 25 MeV, which is only 
visible when applying the cut on low momenta; otherwise  only a slight 
asymmetry in the mass spectrum survives.

We note, however, that the lifetime of the $\phi$ meson at normal 
nuclear matter density might be shorter due to in-medium modifications 
of the kaons and antikaons because for dropping kaon masses the phase 
space for the $\phi$ decay to kaon and antikaon in the medium increases 
\cite{Ko}. While there are presently no strong indications for a shift 
of the kaon mass with density \cite{Brat97c}, some clear evidence for 
an in-medium mass change of antikaons could be established 
\cite{Ko95,Ca97}. The actual calculations of Ref. \cite{Ca97} indicate 
that the antikaon should drop in mass by about 20\% at normal nuclear 
matter density roughly in line with the Lagrangian approaches of Kaplan 
and Nelson \cite{Kaplan} or Waas, Kaiser and Weise \cite{Weise}. As a 
consequence, the 'inside' component of the $\phi$ decay should be 
enhanced in such a scenario. We have also performed studies with a 
modified decay width of the $\phi$ to $K\bar{K}$ as \begin{equation} 
\Gamma_K^* = \Gamma_K^0 \ \{\frac{(M_{\phi}^{*2} -(M_K + 
M_{\bar{K}}^*)^2) (M_{\phi}^{*2} -(M_K - 
M_{\bar{K}}^*)^2)/M_{\phi}^{*2}}{M_{\phi}^2 - 4 M_K^2}\}^{1/2} , 
\end{equation} where $M_{\phi}, \Gamma_K^0$ are the bare $\phi$ mass 
and decay width to $K\bar{K}$, respectively, and $M_{\phi}^*(r)$ is the 
in-medium mass of the $\phi$ that drops with density according to Eq. 
(\ref{Brown}) with a coefficient of 0.025 while the antikaon mass 
$M_{\bar{K}}^*$ drops with density with a coefficient $\approx$ 0.2 
\cite{Ca97}. At normal nuclear matter density the decay width of the 
$\phi$ to $K\bar{K}$ thus increases from 3.8 MeV to $\approx$ 8 MeV by 
roughly a factor of 2 due to the kaonic decay; however, this additional 
width is smaller than the collisional broadening of $\delta \Gamma_R 
\approx$ 15 - 20 MeV such that this effect will be hard to see 
experimentally. We thus discard an explicit representation of this 
limit in Fig. 12 because it almost coincides with the dashed histogram.

\section{Summary}
In this work we have presented fully microscopic calculations for 
dilepton production in pion-nucleus reactions within the INC approach 
and investigated the various contributing channels as well as in-medium 
modifications of the vector mesons due to collisional broadening or 
in-medium mass shifts \cite{1,2,3,4}. Our results for $\pi^-  $ Pb at 
1.3 GeV/c indicate that the dominant background for invariant masses 
$M$ above 0.6 GeV arises from $\pi^- N$ bremsstrahlung which, however, 
is still small compared to the yield from the direct vector meson 
decays.  A mass shift of the $\rho$ and $\omega$ mesons should be seen 
experimentally by an enhanced yield in the mass regime 0.65 $\leq M 
\leq$ 0.75 GeV and a mass shift of the $\rho$ meson especially in a 
decrease of the dilepton yield for $M \geq $ 0.85 GeV because the 
$\rho$ almost entirely decays inside a Pb-nucleus. The in-medium 
modifications of the $\omega$ mesons are found to be most pronounced 
for small momentum cuts on the $e^+e^-$ pair in the laboratory ($P_z 
\leq$ 0.25 GeV/c, $P_T \leq$ 0.25 GeV/c).

We have, furthermore, addressed the production and decay of $\phi$ 
mesons in $\pi^-  A$ reactions. For $\pi^-  Pb$ at 1.7 GeV/c we find a 
sufficient cross section for $\phi$ production; its dilepton decay 
signal is clearly visible above the background from $\rho^0$ decays. 
The in-medium mass shift of the $\phi$ is expected to be much smaller 
than that of the $\rho, \omega$ mesons and the dominant effect expected 
is a broadening of the $\phi$ peak due to elastic $\phi N$ collisions 
and an enhanced kaonic decay width in-medium due to a dropping antikaon 
mass \cite{Ca97,Weise}.  In order to distinguish experimentally the 
in-medium $\phi$ peak from the vacuum decay our analysis indicates that 
sensible cuts on low dilepton momenta in the laboratory as well as an 
experimental mass resolution $\Delta M \leq$ 5 MeV will be necessary.

\vspace{1cm}
The authors acknowledge many helpful discussions with K. Boreskov, E.L. 
Bratkovskaya, H. Bokemeyer, A. Iljinov, W. Koenig, V. Metag, W. 
Sch\"on, A.  Sibirtsev and Yu. Simonov throughout this study.

\newpage
\section*{Figure captions}

\vspace{5mm}
\noindent
{\bf Fig.~1:} The momentum distribution of the $\omega$ mesons produced 
in the reaction $\pi^- + Pb \to \omega X$\ at 1.3 GeV/c as a function 
of the longitudinal momentum in the laboratory $P_z$ (upper part) and 
the transeverse momentum $P_T$ (lower part); solid histograms:   
$\omega$ mesons decaying inside the nucleus, dashed histograms: 
$\omega$ mesons decaying in the vacuum.

\vspace{5mm}
\noindent
{\bf Fig. 2:} The inclusive differential cross section of $e^+e^-$ 
pairs from direct $\omega$ decays for $\pi^- + Pb$ at 1.3 GeV/c for 
different cuts in longitudinal ($P_z$)  and transverse ($P_T$) 
momentum.  The ratio R (\ref{R}) provides a measure for the relative 
weight of the in-component relative to the vacuum component using a cut 
in invariant mass at $M_c$ = 0.725 GeV.

\vspace{5mm}
\noindent
{\bf Fig. 3:} The inclusive differential cross section of $e^+e^-$ 
pairs from direct $\omega$ decays for $\pi^- $ induced reactions at 1.3 
GeV/c on $^{208}Pb, ^{90}Zr, ^{40}Ca$ and $^{12}C$ for a low momentum 
cut ($P_z \leq$ 0.25 GeV/c, $P_T \leq$ 0.25 GeV/c).  The ratio R 
(\ref{R}) provides a measure for the relative weight of the 
in-component relative to the vacuum component using a cut in invariant 
mass at $M_c$ = 0.725 GeV.

\vspace{5mm}
\noindent
{\bf Fig. 4:} The inclusive differential cross section of $e^+e^-$ 
pairs from direct $\omega$ decays for $\pi^- $ induced reactions on 
$^{208}Pb$ at 1.1, 1.3, 1.5 and 1.7 GeV/c  for a low momentum cut ($P_z 
\leq$ 0.25 GeV/c, $P_T \leq$ 0.25 GeV/c).  The ratio R (\ref{R}) 
provides a measure for the relative weight of the in-component relative 
to the vacuum component using a cut in invariant mass at $M_c$ = 0.725 
GeV.

\vspace{5mm}
\noindent
{\bf Fig.~5:} The momentum distribution of the $\omega$ mesons produced 
in the reaction $\pi^-  Pb \to \omega X$\ at 1.3 GeV/c as a function of 
the momentum in the laboratory $P$ for the 'inside' (upper part) and 
the 'outside' component (lower part). The dashed histograms result from 
a calculation using the elastic $\omega N$ cross section (\ref{c5}) 
(b), while the solid histograms reflect a calculation with twice the 
elastic $\omega N$ cross section (a).

\vspace{5mm}
\noindent
{\bf Fig. 6:} The inclusive differential cross section of $e^+e^-$ 
pairs from direct $\omega$ decays for $\pi^- $ induced reactions on 
$^{208}Pb$ at 1.3  GeV/c for a low momentum cut ($P_z \leq$ 0.25 GeV/c, 
$P_T \leq$ 0.25 GeV/c).  The dashed histogram (b) results from a 
calculation using the elastic $\omega N$ cross section (\ref{c5}), 
while the solid histogram (a) reflects a calculation with twice the 
elastic $\omega N$ cross section.

\vspace{5mm}
\noindent
{\bf Fig. 7:} The inclusive differential cross section of $e^+e^-$ 
pairs from direct $\omega$ decays for $\pi^- $ induced reactions on 
$^{208}Pb$ at 1.3 GeV/c for a low momentum cut ($P_z \leq$ 0.25 GeV/c, 
$P_T \leq$ 0.25 GeV/c).  The dashed histogram (b) results from a 
calculation using a $\omega$ mass shift of 18\% at $\rho_0$ in 
(\ref{Brown}), while the solid histogram (a) reflects a calculation 
with half the mass shift (9\%).

\vspace{5mm}
\noindent
{\bf Fig. 8:} The inclusive differential cross section of $e^+e^-$ 
pairs from $\pi^- $ induced reactions on $^{208}Pb$ at 1.3  GeV/c; 
l.h.s.: integrated over all dilepton momenta; r.h.s.: for a low 
momentum cut ($P_z \leq$ 0.25 GeV/c, $P_T \leq$ 0.25 GeV/c). The upper 
solid line represents the sum of all contributions; the thin histograms 
present the yields from pion-nucleon bremsstrahlung ($\pi N$), the 
$\eta$ Dalitz decay ($\eta$), the $\omega$ Dalitz decay and the direct 
decays of the vector mesons $\rho$ and $\omega$, respectively.

\vspace{5mm}
\noindent
{\bf Fig. 9:} Comparison of the inclusive differential cross section of 
$e^+e^-$ pairs from $\pi^- $ induced reactions on $^{208}Pb$ (solid 
histograms) and $^{12}C$ (dashed histograms) at 1.3  GeV/c including 
all sources; l.h.s.: integrated over all dilepton momenta; r.h.s.: for 
a low momentum cut ($P_z \leq$ 0.25 GeV/c, $P_T \leq$ 0.25 GeV/c).  
Both systems have been normalized to the same differential cross 
section in the $\omega$ vacuum decay peaks.

\vspace{5mm}
\noindent
{\bf Fig.~10:} The momentum distribution of the $\phi$ mesons produced 
in the reaction $\pi^-  Pb \to \phi X$\ at 1.7 GeV/c as a function of 
the longitudinal momentum in the laboratory $P_z$ (upper part) and the 
transverse momentum ($P_t$) (lower part) for the 'inside' (solid 
histograms) and the 'outside' components (dashed histograms).

\vspace{5mm}
\noindent
{\bf Fig.~11:} The invariant mass distribution of dilepton pairs from 
$\omega, \rho^0$ and $\phi$ decays for $\pi^-  Pb$ at 1.7 GeV/c 
including the mass shift (\ref{Brown}) for $\omega$'s and $\rho$'s as 
well as collisional broadening for all mesons, however, discarding a 
mass shift of the $\phi$ meson; l.h.s.:  momentum integrated mass 
spectra; r.h.s.: for a low momentum cut ($P_z \leq$ 0.25 GeV/c, $P_T 
\leq$ 0.25 GeV/c).

\vspace{5mm}
\noindent
{\bf Fig.~12:} The invariant mass distribution of dilepton pairs from 
$\omega, \rho^0$ and $\phi$ decays for $\pi^-  Pb$ at 1.7 GeV/c 
including the mass shift (\ref{Brown}) for $\omega$'s and $\rho$'s as 
well as collisional broadening for these mesons for a low   momentum 
cut ($P_z \leq$ 0.25 GeV/c, $P_T \leq$ 0.25 GeV/c).  The full histogram 
at $\approx$ 1.02 GeV of invariant mass shows the $\phi$ decay without 
a mass shift but collisional broadening (cf. Fig. 11) while the dashed 
histogram additionally includes a $\phi$ mass shift according to 
(\ref{Brown}) by 2.5\% at $\rho_0$.

\newpage
\psfig{figure=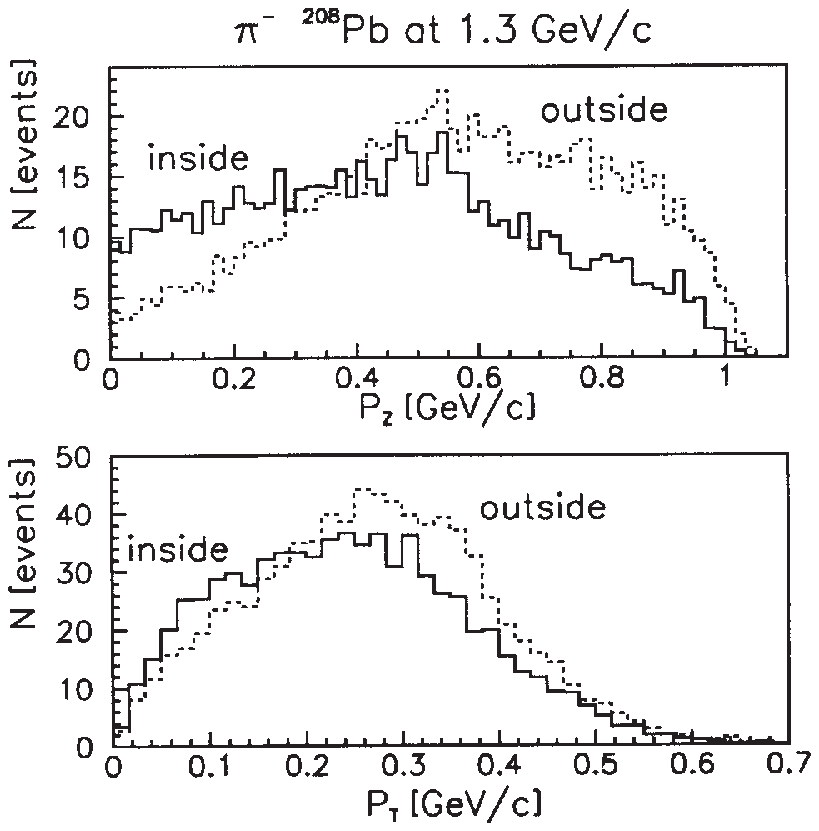,width=16cm,height=23cm,angle=270}
\vspace*{-3cm}
Fig. 1
\newpage
\psfig{figure=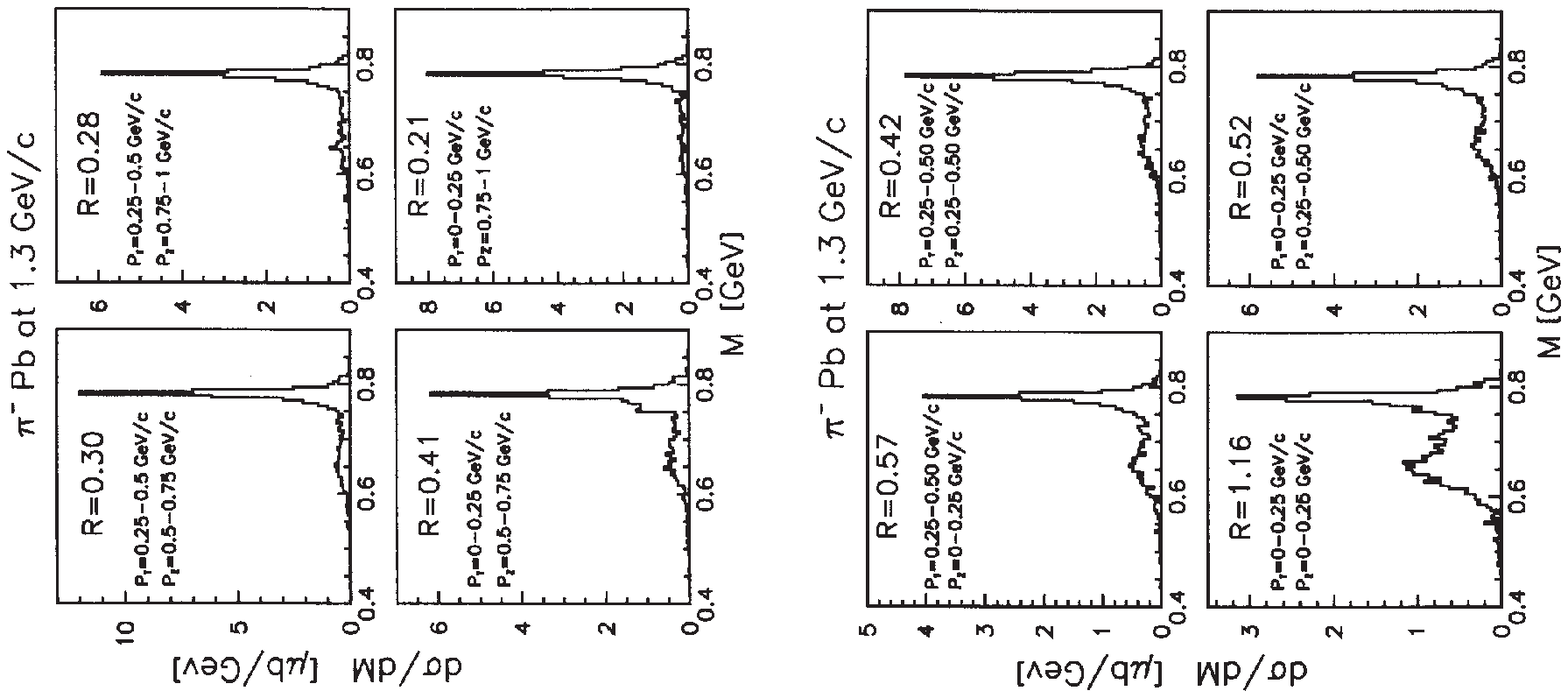,width=16cm,height=23cm,angle=270}
\vspace*{-1cm}
Fig. 2
\newpage
\psfig{figure=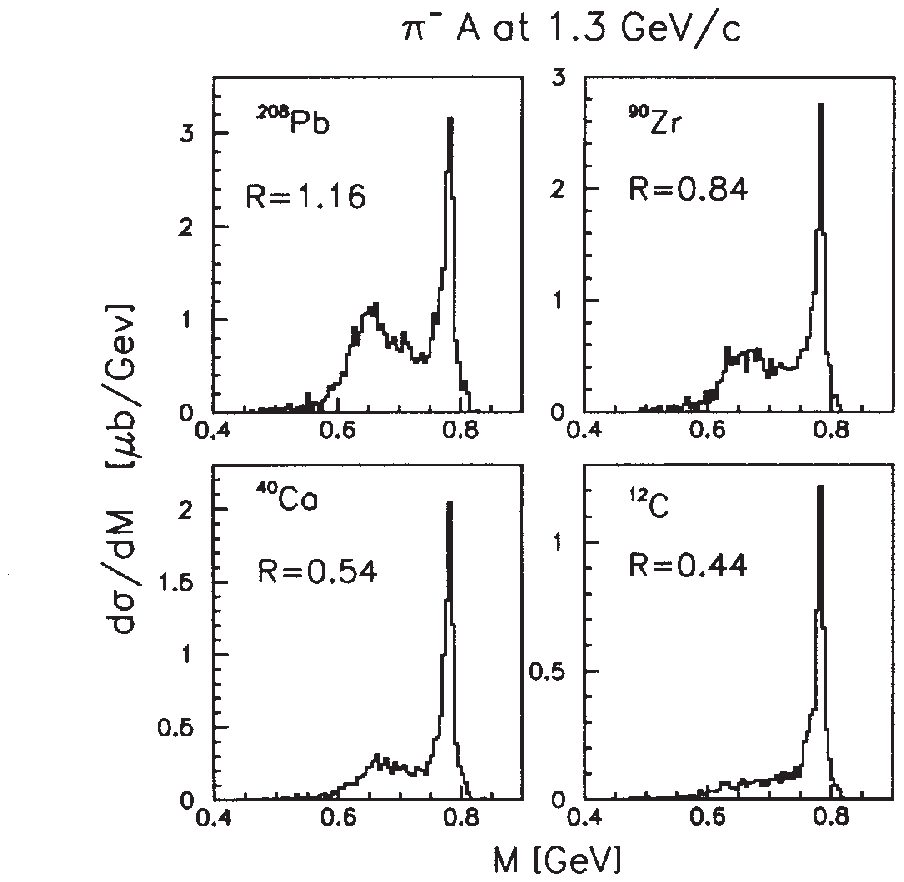,width=16cm,height=23cm,angle=270}
\vspace*{-1cm}
Fig. 3
\newpage
\psfig{figure=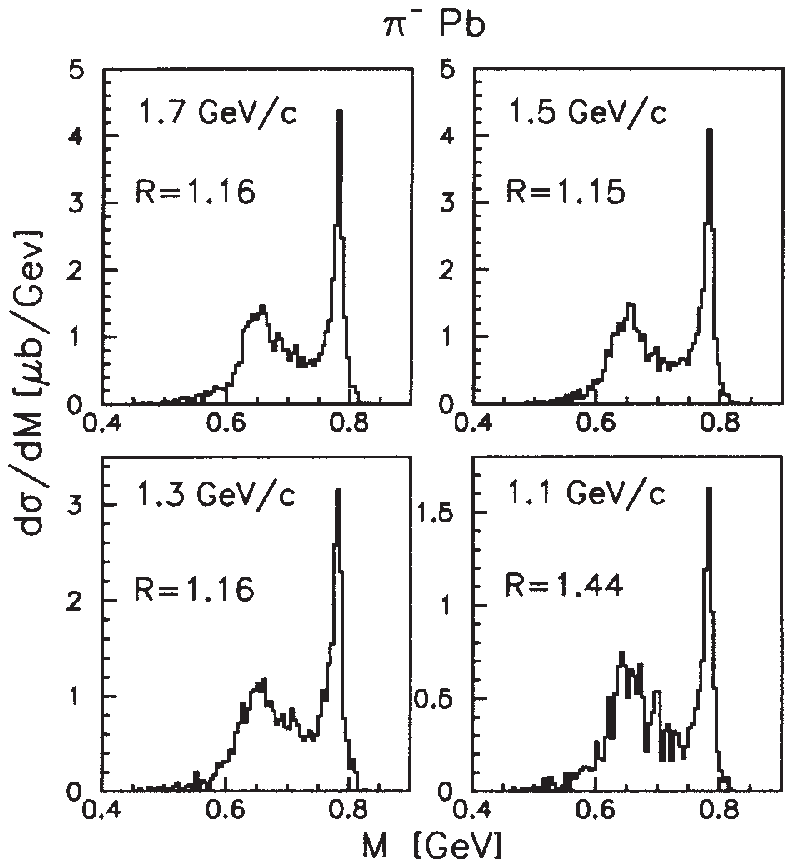,width=16cm,height=23cm,angle=270}
\vspace*{-3cm}
Fig. 4
\newpage
\psfig{figure=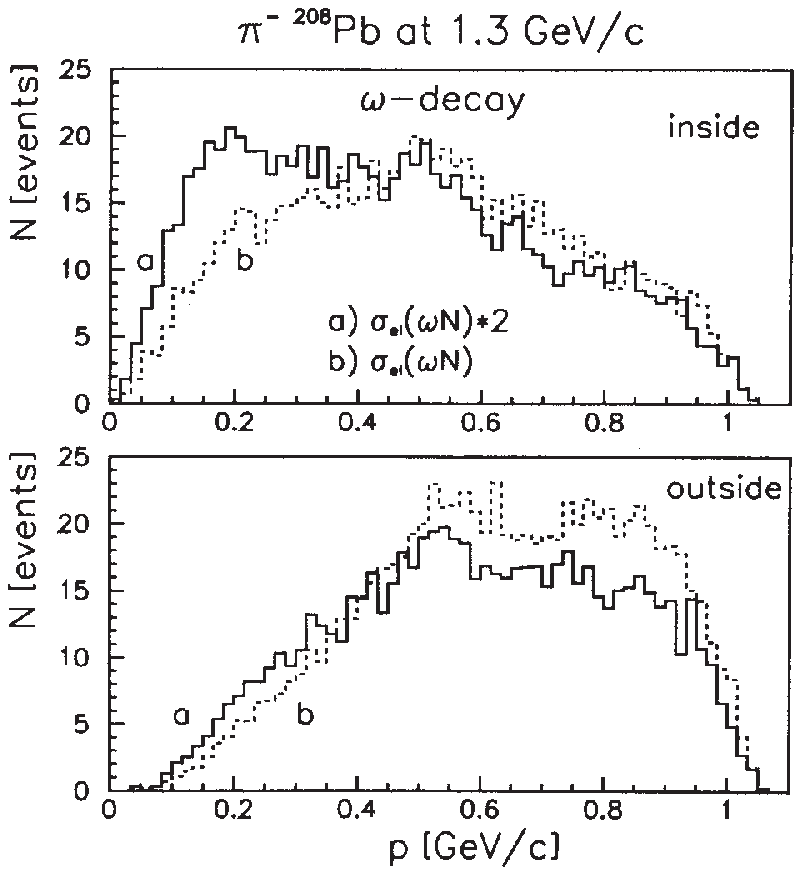,width=16cm,height=23cm,angle=270}
\vspace*{-3cm}
Fig. 5
\newpage
\psfig{figure=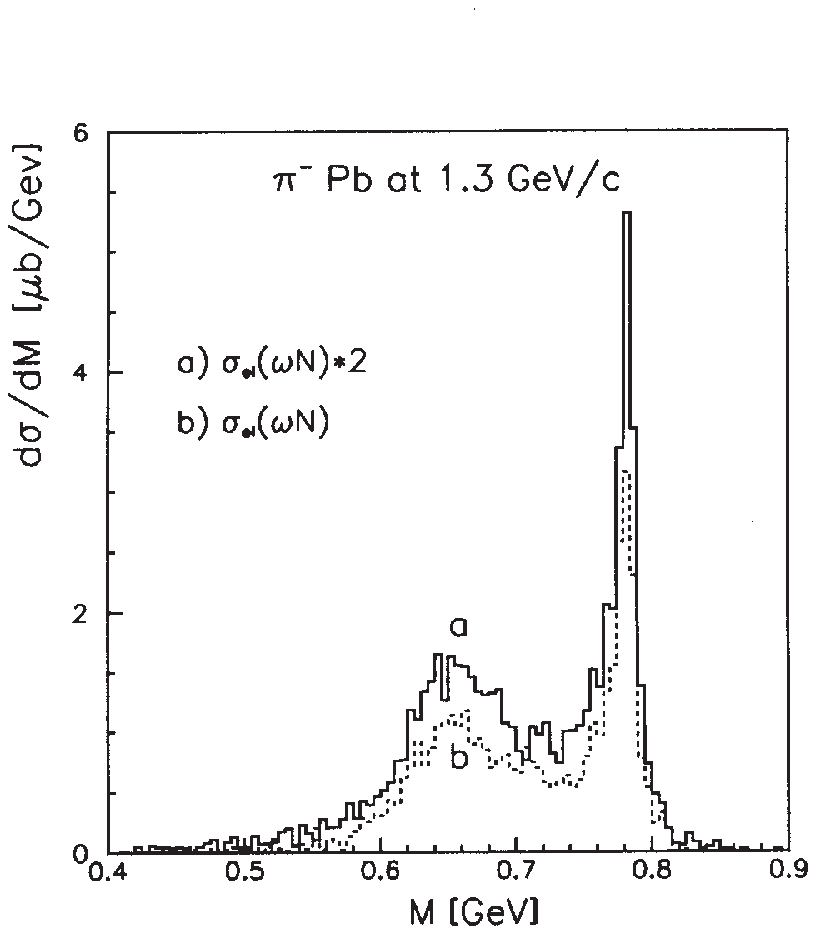,width=16cm,height=23cm,angle=270}
\vspace*{-3cm}
Fig. 6
\newpage
\psfig{figure=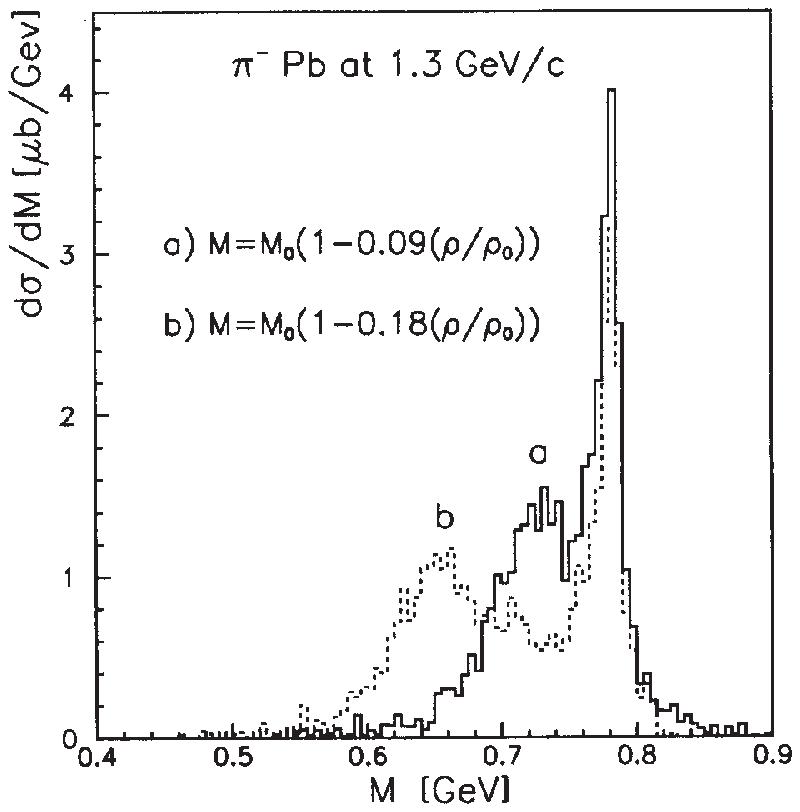,width=16cm,height=23cm,angle=270}
\vspace*{-3cm}
Fig. 7
\newpage
\psfig{figure=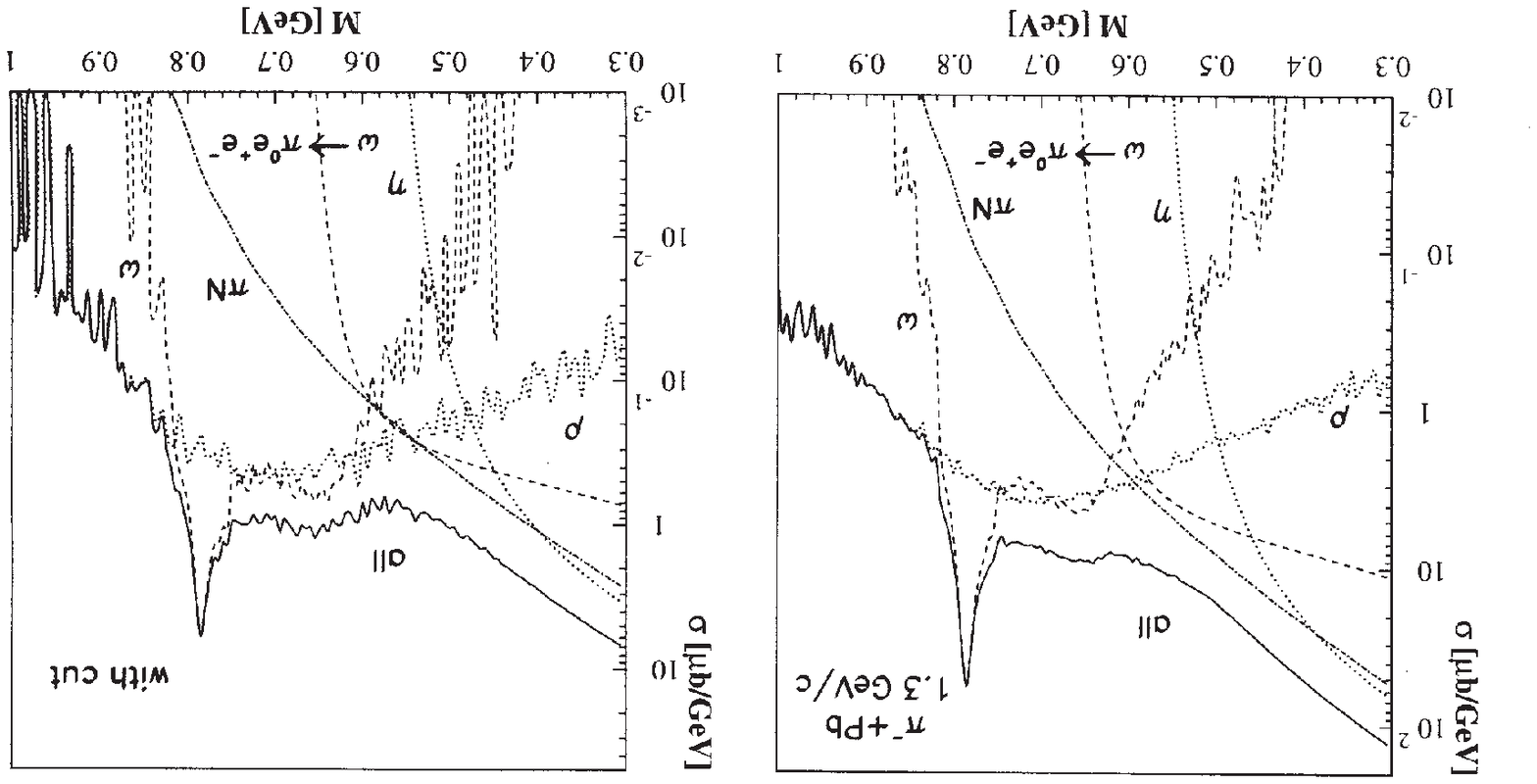,width=16cm,height=23cm,angle=270}
\vspace*{-1cm}
Fig. 8
\newpage
\psfig{figure=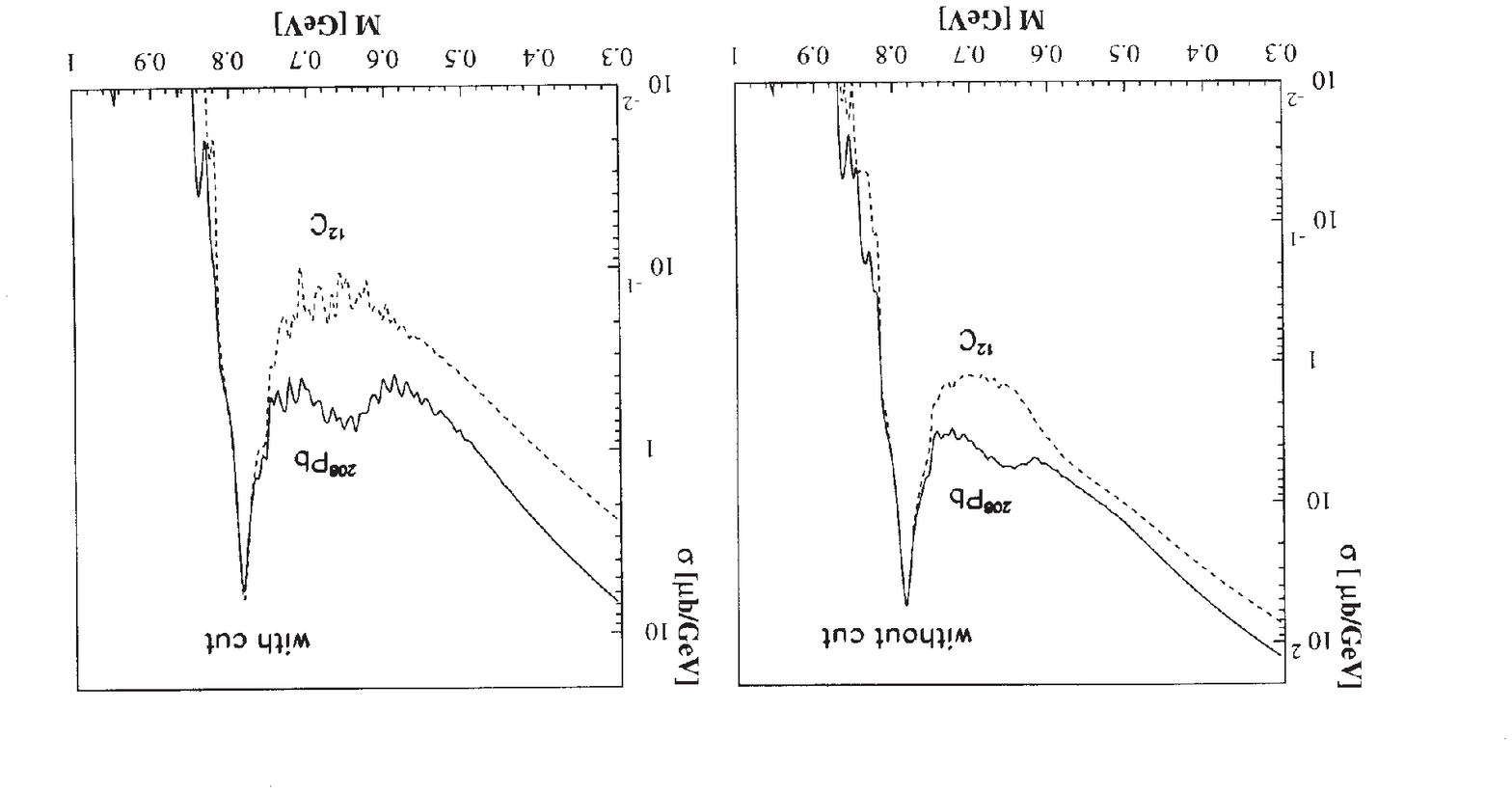,width=16cm,height=23cm,angle=270}
\vspace*{-1cm}
Fig. 9
\newpage
\psfig{figure=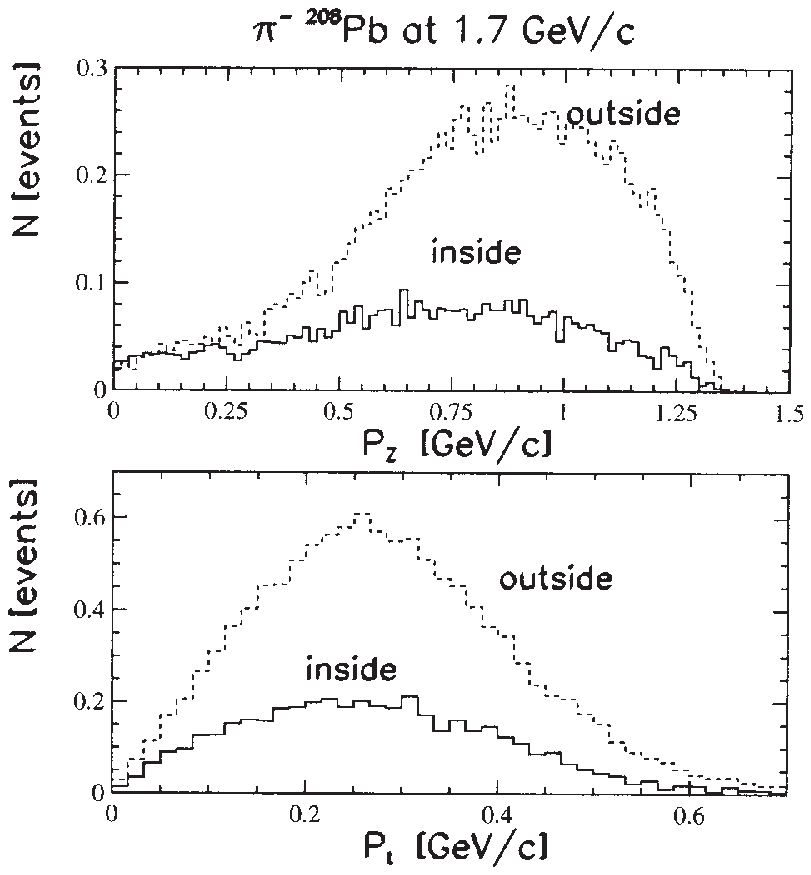,width=16cm,height=23cm,angle=270}
\vspace*{-1cm}
Fig. 10
\newpage
\psfig{figure=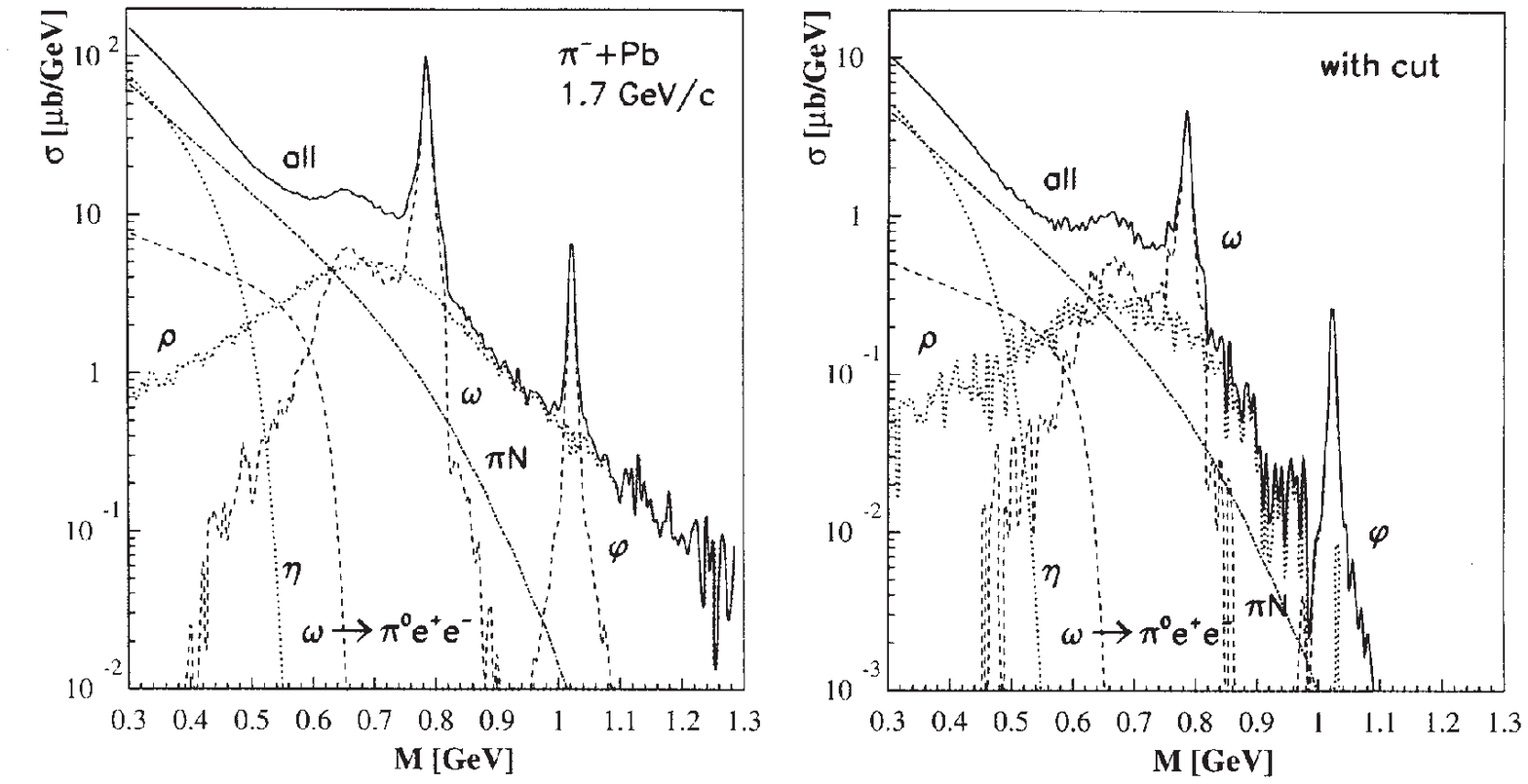,width=16cm,height=23cm,angle=90}
\vspace*{-1cm}
Fig.11
\newpage
\psfig{figure=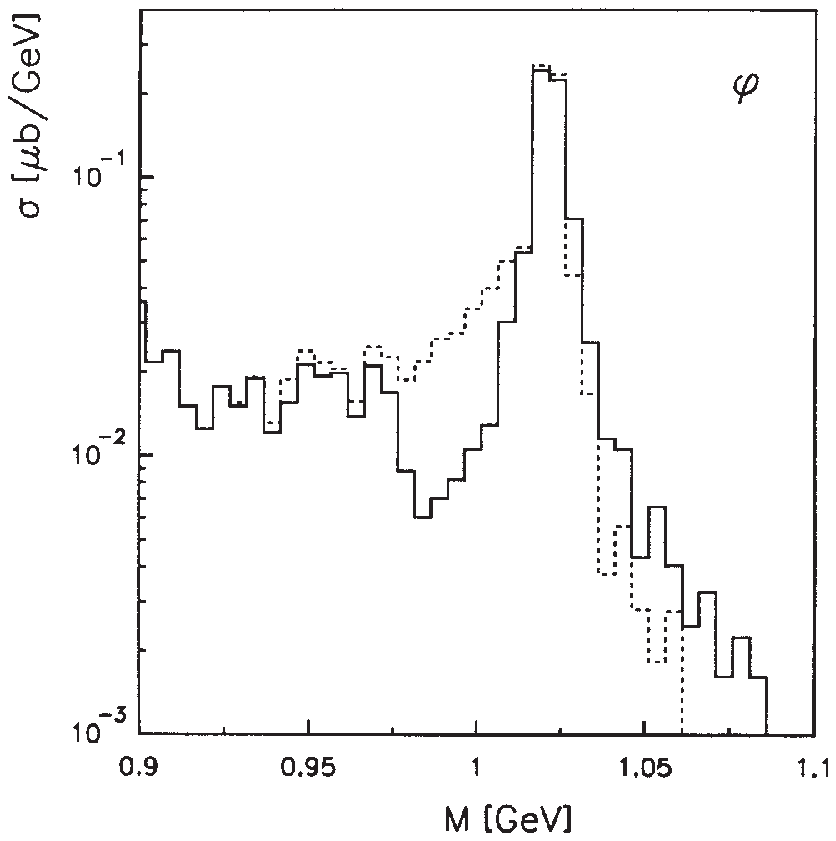,width=16cm,height=23cm,angle=270}
\vspace*{-3cm}
Fig.12

\end{document}